\documentclass[journal=jacsat,manuscript=article]{achemso}

\usepackage[english]{babel}
\usepackage{amssymb,amsmath}
\usepackage{graphicx}

\author{L\'{e}onard Desvignes}
\affiliation[LPS]
{Université Paris-Saclay, CNRS, Laboratoire de Physique des Solides, 91405, Orsay, France}
\author{Vasily S. Stolyarov}
\affiliation[Moscow]
{Moscow Institute of Physics and Technology, 141700 Dolgoprudny, Russia}
\author{Marco Aprili}
\author{Freek Massee}
\email{freek.massee@universite-paris-saclay.fr}
\affiliation[LPS]
{Université Paris-Saclay, CNRS, Laboratoire de Physique des Solides, 91405, Orsay, France}

\title[Tunable high speed atomic rotor in Bi$_2$Se$_3$ revealed by current noise]
{Tunable high speed atomic rotor in Bi$_2$Se$_3$ revealed by current noise}

\keywords{atomic rotor, scanning tunnelling microscopy/spectroscopy, current noise, finite frequency, Bi$_2$Se$_3$}

\begin{document}

\begin{abstract}
The ability to manipulate individual atoms and molecules using a scanning tunnelling microscope (STM) has been crucial for the development of a vast array of atomic scale devices and structures ranging from nanoscale motors and switches to quantum corrals. Molecular motors in particular have attracted considerable attention in view of their potential for assembly into complex nanoscale machines. Whereas the manipulated atoms or molecules are usually on top of a substrate, motors embedded in a lattice can be very beneficial for bottom-up construction, and may additionally be used to probe the influence of the lattice on the electronic properties of the host material. Here, we present the discovery of controlled manipulation of a rotor in Fe doped Bi$_2$Se$_3$. We find that the current into the rotor, which can be finely tuned with the voltage, drives omni-directional switching between three equivalent orientations, each of which can be frozen in at small bias voltage. Using current fluctuation measurements at 1\thinspace MHz and model simulations, we estimate that switching rates of hundreds of kHz for sub-nA currents are achieved.
\end{abstract}

Manipulation of single atoms and molecules\cite{foster_nature_1988, eigler_nature_1990} has been achieved through mechanical contact \cite{bartels_prl_1997, ternes_science_2008}, the local electric field\cite{rezaei_jcp_1999, alemani_jacs_2006, massee_science_2020} and the tunnelling current\cite{mo_science_1993, fishlock_nature_2000, komeda_science_2002, sloan_2011}, leading to a wealth of artificial devices and structures\cite{eigler_nature_1991, gimzewski_science_1998, manoharan_nature_2000, stroscio_science_2006, grill_naturenano_2007, moon_science_2008}. Manipulation using the tunnelling current is especially interesting as the current is highly localised due to its exponential dependence on the tip-sample distance, as well as tunable and non-invasive. However, unlike mechanical- and electric field manipulations, reading and writing using the tunnelling current is difficult to perform independently as these operations often occur for similar parameters. Ideally, therefore, one would like to have an extra control parameter to tune what fraction of the total current flows into the atom or molecule: a low fraction for reading and high fraction for writing. Here we show how the voltage can be used to finely tune the current flowing into an embedded rotor we discovered in Fe-doped Bi$_2$Se$_3$, thereby allowing precise control over the rotor activation.

\section{Results and Discussion}
\begin{figure}
	\centering
	\includegraphics[width=3.33in]{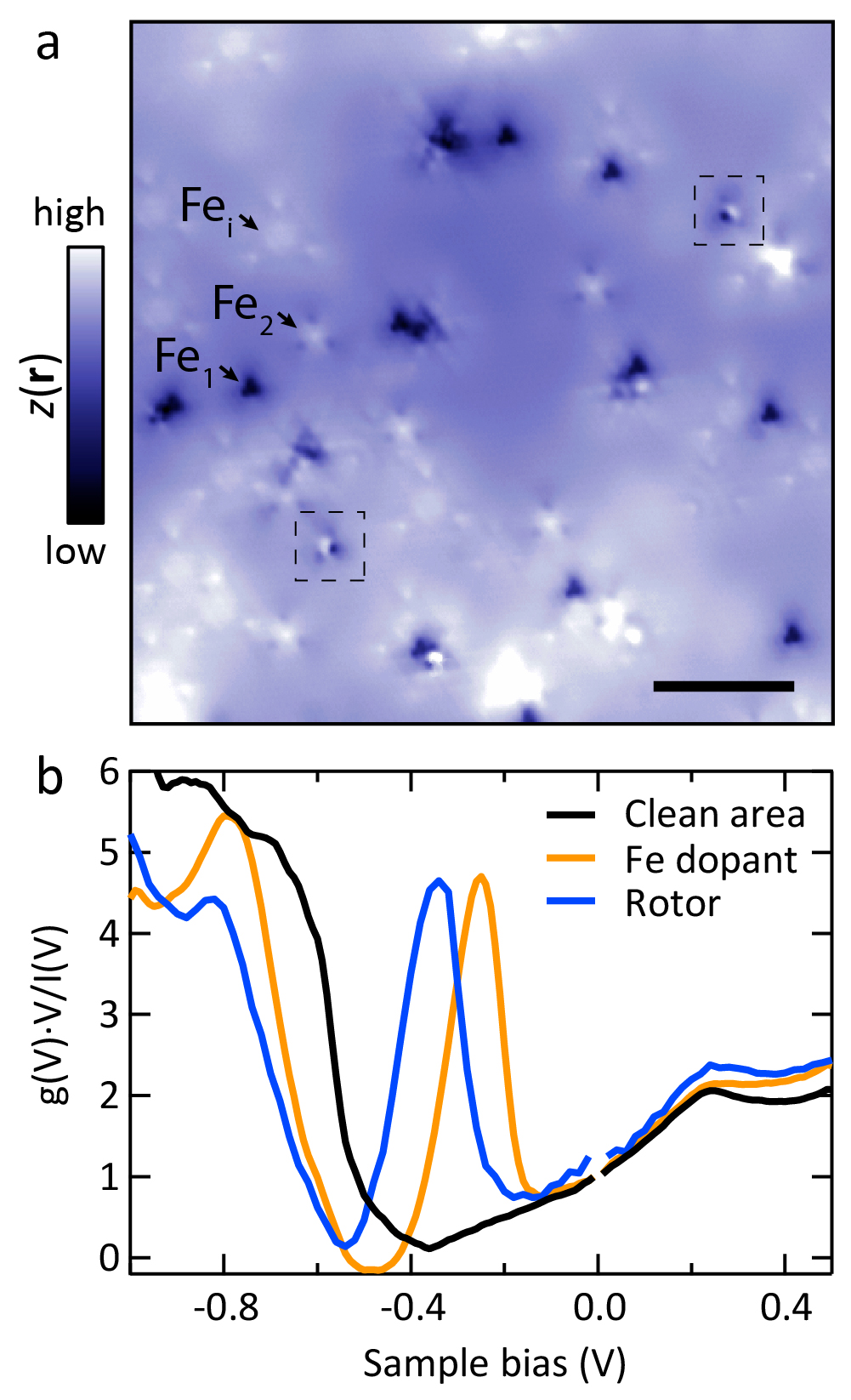}
	\caption{\label{fig:1} Symmetry breaking defects in Bi$_2$Se$_3$. \textbf{a} Constant current image of 0.2 \% Fe doped Bi$_2$Se$_3$. Fe impurities replacing a Bi atom in the first and second Bi layer below the surface are indicated (Fe$_1$ and Fe$_2$, respectively), as well as an interstitial Fe impurity (Fe$_{i}$). The two objects in dashed boxes do not preserve the triangular symmetry of the lattice. $V_{\text{bias}} =$ 100\thinspace mV, $I =$ 100\thinspace pA, the scale bar is 5\thinspace nm. \textbf{b} Normalised differential conductance spectra on a clean area, on an Fe impurity in the first layer below the surface and on a symmetry breaking object. $V_{\text{set}} =$ -1\thinspace V, $I_{\text{set}} =$ 400\thinspace pA.}.
\end{figure}

Figure \ref{fig:1}a shows a constant current image of the selenium top-layer of the Bi$_2$Se$_3$ which is bulk-doped with 0.2\% Fe. The Fe impurities preferentially replace sub-surface bismuth atoms\cite{zhang_prl_2012, abdalla_prb_2013} and can be identified by the shape and voltage dependence of their localised energy levels\cite{song_prb_2012, stolyarov_apl_2017}. Additionally, due to the smaller Fe-Se bond length\cite{abdalla_prb_2013}, replacement of a Bi atom by an Fe impurity in the first layer below the surface leads to a reduced height of the three neighbouring Se atoms, seen as dark triangles in topography at voltages near the chemical potential, $\epsilon_{F}$. Importantly, all resonant levels and topographic features normally maintain the triangular symmetry of the lattice. Surprisingly, though, we observe a previously unreported object of which two can be seen in Fig. \ref{fig:1}a that does not preserve this symmetry. The size, position and resonant level (see Fig. \ref{fig:1}b) of the object are very similar to that of an Fe impurity in the first layer below the surface, yet only one of the Se atoms is dark, whereas the other two appear brighter. 

On closer inspection, the orientation of the symmetry-breaking object is not uniform across the sample, as illustrated by the two in Fig. \ref{fig:1}a that point in different directions. In fact, we can rotate each object independently, evidenced by the three orientations of a single 'rotor' shown in Figure \ref{fig:2}a-c, and the large field of view images in Supplementary Information section 1. To activate the rotor, we position our tip above it and lower the bias voltage at a fixed average current. Once the sample bias is sufficiently negative ($V_{\text{bias}}<$ -0.16\thinspace V), the current starts to fluctuate between three values corresponding to the three orientations, see Fig. \ref{fig:2}d. Both clock-wise and anti-clockwise manipulations are observed, suggesting the rotor does not have a preferred direction of rotation. When we tune the voltage closer to the peak of the resonance of the rotor at a fixed current, the switching rate increases rapidly, see Fig. \ref{fig:2}e, and already becomes too fast to track within a span of tens of millivolt. This suggests that the current directly into the rotor is driving the rotation. The behaviour at positive sample bias, where in absence of a strong, localised resonance the rotation activation is significantly weaker supports this notion. Importantly, when we change the tip-sample distance at a fixed voltage, the activation rate shows a linear dependence on the current (Fig. \ref{fig:2}f): a hallmark of current-induced activation\cite{sloan_2011}. 

\begin{figure}
	\centering
	\includegraphics[width=\textwidth]{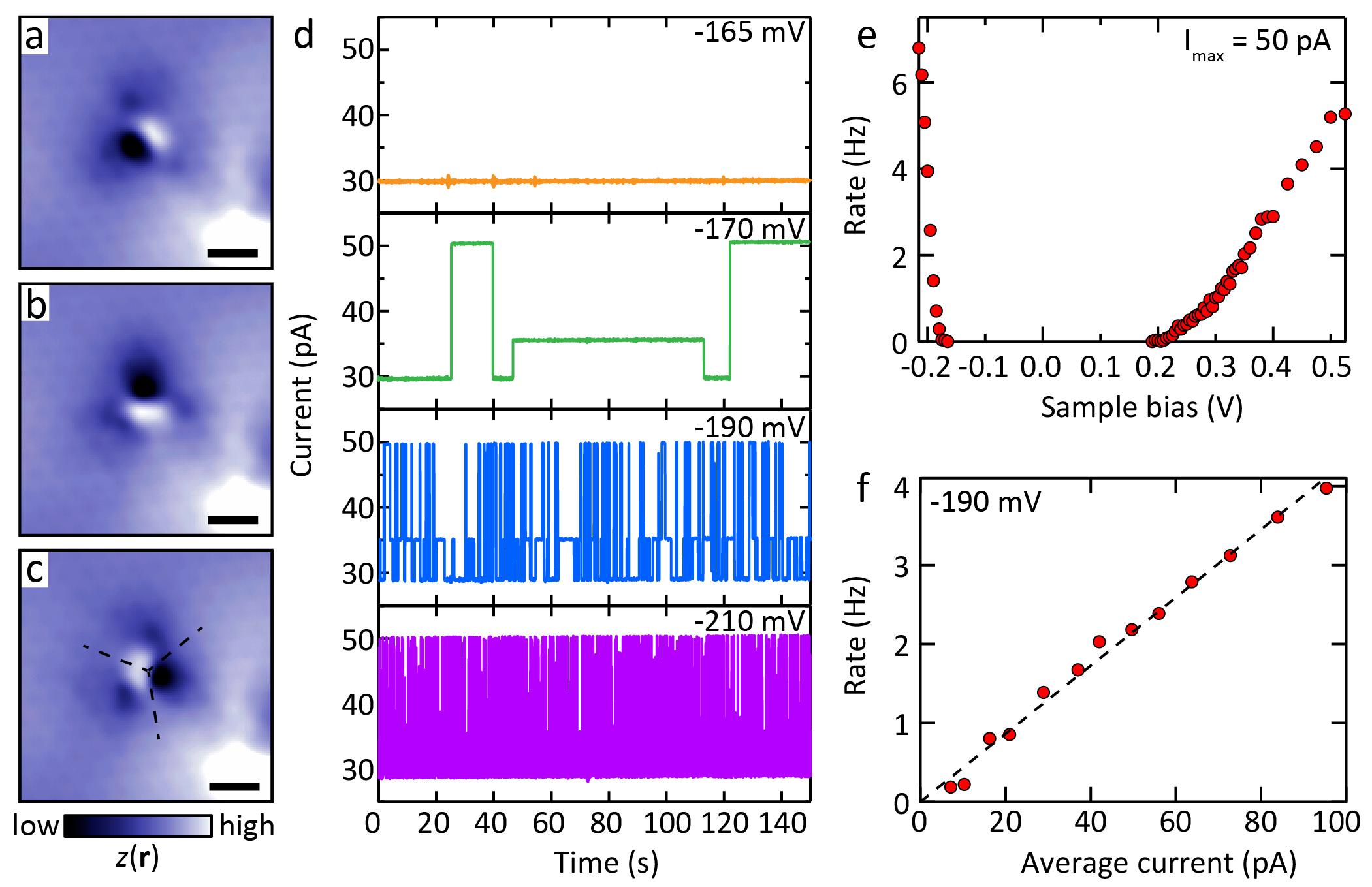}
	\caption{\label{fig:2} Rotor activation. \textbf{a}-\textbf{c} Three orientations of the rotor in the top-right of Fig. \ref{fig:1}a. $V_{\text{bias}} =$ 100\thinspace mV, $I =$ 100\thinspace pA, the scale bar is 1\thinspace nm, the three high symmetry lines where two orientations have the same height are indicated in \textbf{c}. \textbf{d} Current as function of time for different voltages at the same xy-location; the tip-sample distance for each voltage is set such that the current fluctuates in the same range. \textbf{e} Switching rate as function of voltage, including the traces in \textbf{d}. The maximum current is 50\thinspace pA for all points. The rate rapidly increases upon entering the resonance of the rotor at negative voltages (see Fig. \ref{fig:1}b), whereas the rate never becomes too fast to track at positive bias. \textbf{f} Current dependence of the switching rate at a fixed voltage of -190\thinspace mV, the dashed line is a linear fit.}
\end{figure}

\begin{figure}
	\centering
	\includegraphics[width=3.33in]{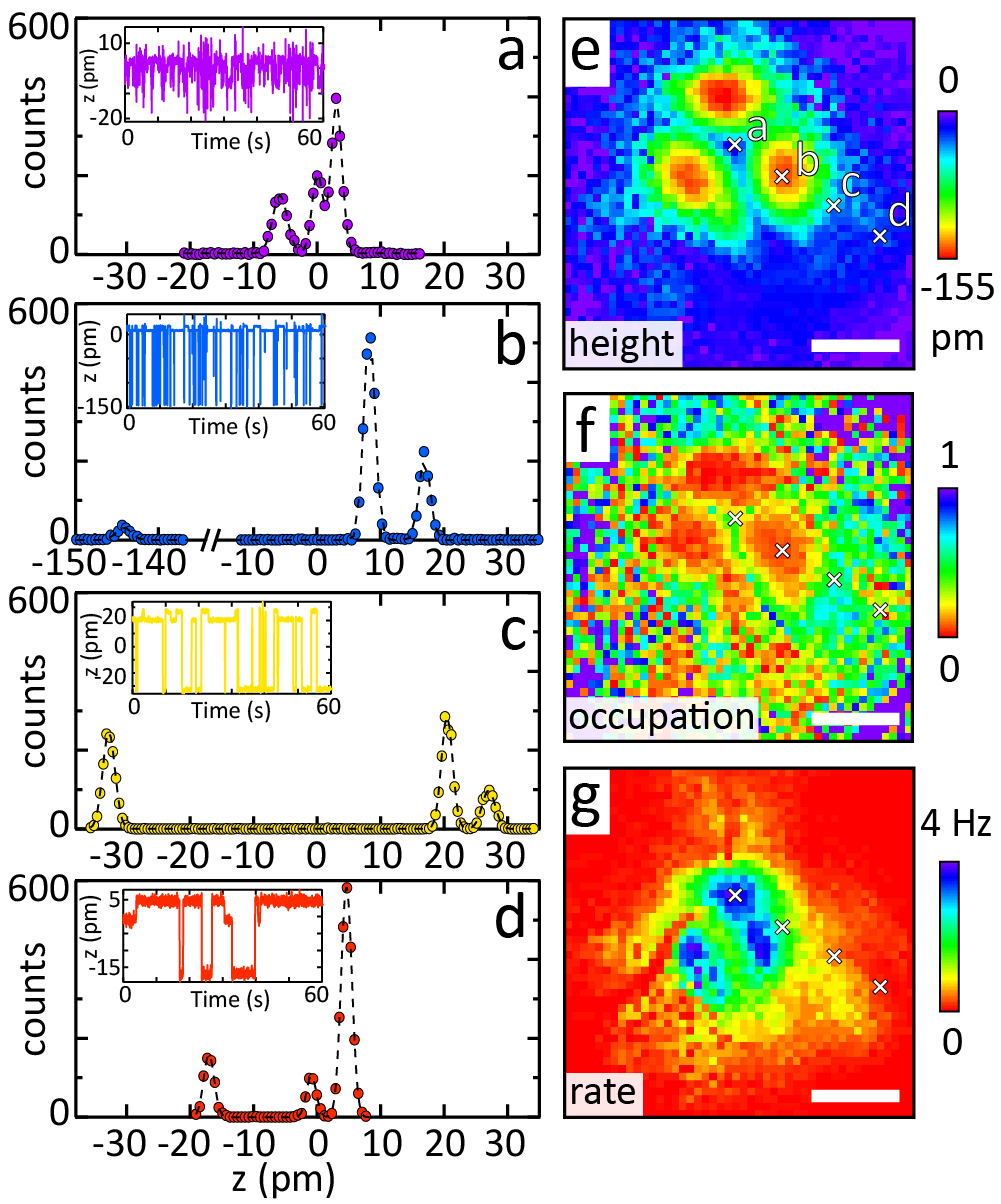}
	
	\caption{\label{fig:3} Spatial dependence of rotation. \textbf{a}-\textbf{d} Histograms of the height fluctuations for 60 second time traces (insets) at the four locations marked in (e). Three Gaussians are fitted (dashed lines) to extract the position and area of each height (i.e. orientation). The time traces are taken at a constant current of $I =$ 50\thinspace pA at $V_{\text{bias}} = $ -0.195\thinspace V. The average height for each time trace is set to zero. \textbf{e} Average height of the lowest value (i.e. left-most peak position in histograms such as a-d), the minima (red colour) correspond to the sites of the depression in Fig. \ref{fig:2}a-c. \textbf{f} Occupation of the lowest value. An occupation of 1 is reached when all three values are identical in height and/or in absence of rotation. \textbf{g} Activation rate on the same field of view as e-f. The highest rate is found near the centre of the rotor. The crosses in f, g are the same as those in e. The scale bar in e-g indicates 0.5\thinspace nm.}
\end{figure}

Having established that we can controllably rotate an object in Bi$_2$Se$_3$, we now address the spatial dependence of the rotation activation. To do so, we record the tip height as function of time and position at a fixed voltage and current. In general, three heights can be observed, corresponding to the three orientations of the rotor. Only on the high symmetry lines (see Fig. \ref{fig:2}c) two orientations will have the same height, while at the centre of the rotor all three do so. We extract the height of the three orientations as well as their relative occupation ratio by fitting Gaussian peaks to the histogram of each time trace, see Fig. \ref{fig:3}a-d. Figure \ref{fig:3}e shows the extracted height of the lowest value, where the minima correspond to the depression in topography for the three orientations as seen in Fig. \ref{fig:2}a-c. Interestingly, exactly at these minima, the lowest value itself is least occupied of the three (Fig. \ref{fig:3}b and f). In other words, the orientation of the rotor that places the depressed (dark) Se atom directly underneath the tip is less favourable. Lastly, Fig. \ref{fig:3}g displays the activation,  

or switching, rate extracted from the same dataset. The rate is highest at the centre of the rotor, and somewhat asymmetric with respect to the centre. We note here that two factors complicate the extraction of the switching rate accurately: 1) our acquisition rate is limited to 0.02 points per second, and 2) the three orientations of the rotor cannot be clearly distinguished near high symmetry lines. Both factors will lead to an underestimation of the actual rate.

\begin{figure}
	\centering
	\includegraphics[width=\textwidth]{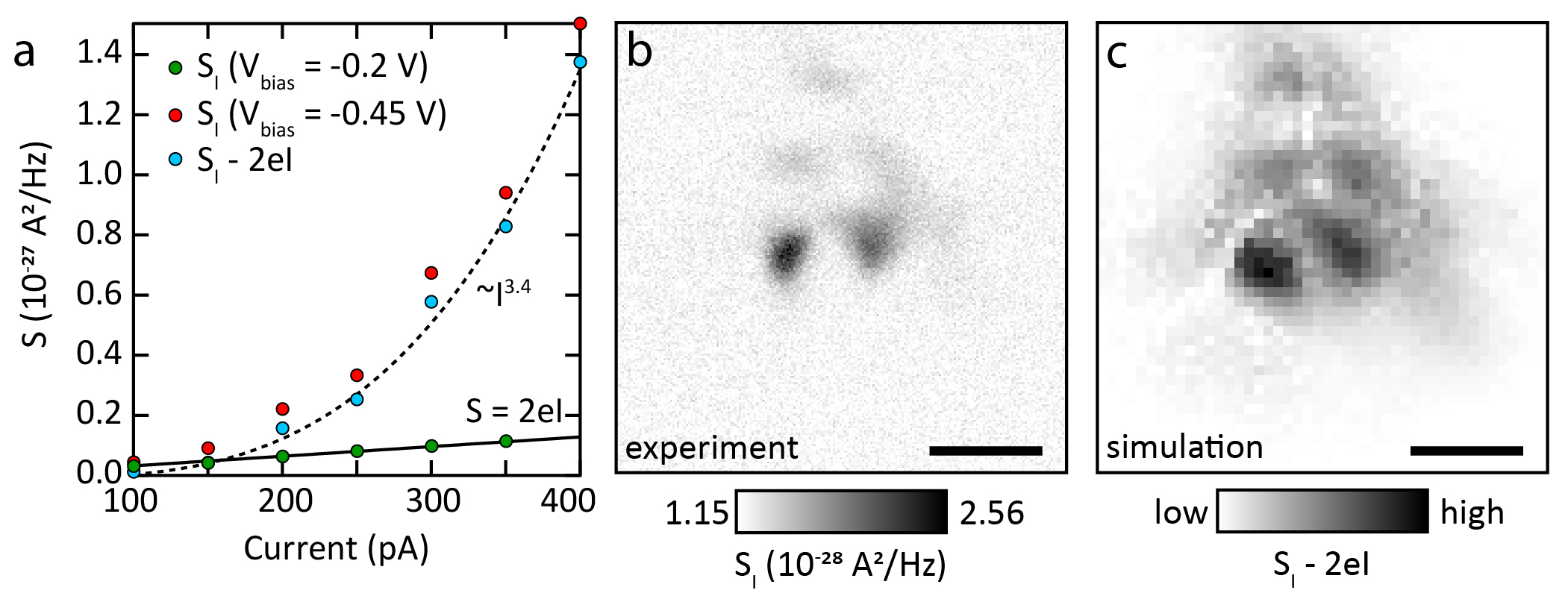}
	\caption{\label{fig:4} Current noise at 1\thinspace MHz. \textbf{a} Current dependence of the current noise: for low switching rates S$_I$ = 2eI (green markers), whereas for high switching rates additional current noise appears (red markers). The excess noise, S$_I$ - S$_{\text{shot}}$, is fitted with a power law (blue dots and dashed line). \textbf{b} Spatially resolved current noise at -0.3\thinspace V and 400\thinspace pA, see Supplementary Information section 4 for additional voltages. \textbf{c} Simulated current noise using the parameters extracted from the experimental data in Fig. \ref{fig:3}. The scale bar in b-c indicates 0.5\thinspace nm.}
\end{figure}

To determine the activation rate at elevated currents, we can circumvent our limited bandwidth by using a parallel circuit we have installed to measure current noise at 1\thinspace MHz\cite{revsciinstrum_massee}. In absence of current fluctuations from the rotor, the total current noise at this elevated frequency, $S_{\text{tot}}$, is equal to the sum of the thermal noise of the circuitry, background noise of e.g. amplifiers, and the shot-noise originating from the charge carriers traversing the vacuum barrier of the junction, $S_{\text{shot}}$. For random tunnelling of electrons, $S_{\text{shot}} = 2eI$, where $e$ is the electron charge and $I$ the tunnelling current\cite{blanter_review_2000}. When we correct for the thermal- and background noise by subtracting the noise at $I = 0$, $S_{0}$, we indeed find that for a stationary rotor the remaining current noise is purely due to shot-noise, $S_I = S_{\text{tot}} - S_{0} = 2eI$ (see Fig. \ref{fig:4}a). Upon activating the rotor, the additional noise at 1\thinspace MHz is negligible for low rotation rates. However, for increasing rates, particularly those for which we can no longer track individual switching events accurately in our real-time signal, $S_{I} > 2eI$. The excess noise introduced by the rotor is extracted by subtracting the shot-noise from $S_I$ and can be fitted with a power law for which we generally obtain an exponent on the order of 3. The spatial profile of $S_{I}$ is shown in Fig. \ref{fig:4}b for the same rotor as Fig. \ref{fig:3}, but at a significantly higher switching rate as a result of choosing a more negative voltage and higher current.

To verify that rotation is causing the excess noise in Fig. \ref{fig:4}b, and subsequently convert our noise data to rotation rates, we construct a simple model to simulate the rotation induced current noise at 1\thinspace MHz. First, we generate time-traces such as shown in Fig. \ref{fig:2} using as input the current of the three orientations of the rotor, $I_i$ with i = 1, 2, 3, their relative occupation, $n_i$ such that $\sum_{i} n_i = 1$, and the total number of switches per time unit. Within these constraints, the duration of each orientation, and rotation direction, are chosen randomly. For simplicity we assume that the jump in current between two orientations is instantaneous and that the $n_i$ are current independent (see Supplementary Information section 3 for more details). By design, our model reproduces the data in Fig. \ref{fig:3} if we use the experimentally determined (low frequency) $n_{i}$ and switching rate, as well as the $I_i$ obtained from converting the three heights at constant current using the measured height dependence of the current (see Supplementary Information section 2). Importantly, since we can choose an arbitrarily large point density, we can also extract the noise at 1\thinspace MHz from the computed power spectral density of the time-traces. The simulated noise, shown in Fig. \ref{fig:4}c, is remarkably similar to the measured noise of Fig. \ref{fig:4}b. This strongly suggests that the excess current noise is indeed exclusively due to an increase in the rotation rate of the rotor. Furthermore, using the linear relation between the switching rate and the average current (\ref{fig:2}f) and exploiting the general properties of a power spectral density, we find that the current noise at a fixed frequency resulting from the switching should follow a cubic dependence on the average current, as seen in experiment (Fig. \ref{fig:4}a, see Supplementary Information section 3 for more details).

Consequently, we can use our model to determine the rotation rate from the current noise measurements. To do so, using the experimentally determined $n_{i}$, we set the $I_i$ such that the average current in the simulation, $\overline{I} = \sum_{i} n_i I_i$, is equal to the experimentally used current, leaving the switching rate as only remaining adjustable parameter. The largest excess noise in Fig. \ref{fig:4}b then requires a rate of $\sim$30\thinspace kHz. For the curve shown in Fig. \ref{fig:4}a, where the bias voltage is set to include the full width of the resonance of the rotor (V$_{\text{bias}}$=-0.45\thinspace V, see Fig. \ref{fig:1}b), the current noise at 400\thinspace pA reaches an absolute maximum of $\sim$ 1.3$\cdot$ 10$^{-27}$\thinspace A$^2$/Hz. From our simulation, we find that this corresponds to a rate of several hundreds of kHz (see Supplementary Information section 3).

Two questions remain to be addressed: what is actually rotating, and why does the rotor prefer not to have its lowest orientation underneath the tip. Constant current imaging suggests that the three surface Se atoms are buckled such that one is below and two above the Se plane (see also Supplementary Information section 1). Based on the size, shape and differential conductance, the object directly below the buckled Se atoms is an Fe impurity. One possibility could be that whereas normally an Fe impurity sits on the Bi site centred in between the Se atoms\cite{zhang_prl_2012, abdalla_prb_2013}, the Fe atom of the rotor is for some reason located off-centre - for example due to a neighbouring impurity or defect. The shift of its resonance with respect to conventional Fe impurities (see Fig.\ref{fig:1}b) can be expected for such a modified environment\cite{stolyarov_apl_2017}. For an off-centre Fe impurity, the Se atom furthest from the Fe impurity will drop below the Se plane, whereas the two Se atoms closer to the Fe impurity will be lifted above the Se plane. The three orientations in this scenario then correspond to the three equivalent positions of the off-centre Fe impurity, where the current into the Fe provides the energy required to hop from one position to another. The spatial dependence of the switching rate naturally follows from spatial variations of the current path into the rotor. Alternatively, a similar Se buckling may occur if the Fe impurity happens to be located slightly above the Bi plane, although it is unclear why activation would be current dependent as the Fe impurity in that case is stationary.

The spatial variation in occupation ratio, however, is more complicated. The low occupation ratio for having the depressed Se atom directly below the tip means that the rotor avoids this configuration and/or that if it does happen, it quickly switches back to another orientation. Our data suggests both mechanisms are relevant. Not only the magnitude of the current is thus important, but also its direction with respect to the rotor orientation. Interestingly, the occupation ratios at positive sample bias are much more homogeneous (see Supplementary Information section 5). This seems to suggest that electrostatics may be important, although more detailed calculations are required to provide a deeper insight into the current polarity and direction dependence of the rotation activation.

\section{Conclusion}
The atomically sized defect embedded in the crystal structure of Bi$_2$Se$_3$ we describe here allows controllable rotation of hundreds of kHz for sub-nA currents. The rotation activation, which we show to be resulting from the tunnelling current directly into the rotor, can be finely regulated by taking advantage of the voltage dependence of the tunnelling process. The current-direction dependence of the occupation ratio additionally allows control over which states are predominantly occupied. The highly tunable rotor may be exploited as nano-sized current sensor, or as starting point for more complicated bottom-up engineered atomic scale structures. Additionally, the use of current noise measurements to analyse high activation rates we introduce in this work can be widely applied in atomic scale structures and opens a new avenue for device characterization.

\section{Methods}
Bi$_{2}$Se$_{3}$ single crystals incorporating 0.2\% Fe-impurities were mechanically cleaved in cryogenic vacuum at T $\sim$ 20\thinspace K and directly inserted into the STM head at 4.2 K. A mechanically cut Pt/Ir tip was used, with energy independent density of states. Differential conductance measurements throughout used a standard lock-in amplifier with a modulation frequency of 429.7Hz. STM and simultaneous noise measurements were performed with the home-built setup and MHz circuitry described in Ref.~\citenum{revsciinstrum_massee}. The noise amplitude spectral density was measured at the LC$_{\text{cable}}$ resonance of 1\thinspace MHz. All measurements were performed at T $<$ 2\thinspace K.

\section{Associated content}
\subsection{Supporting Information}
More details regarding the independent rotor activation, the height to current conversion, the model, the voltage dependence of current noise and the rotation activation at positive bias voltage.

\section{Author Information}

\subsection{Author contributions}
LD and FM performed the measurements and analysed the data. LD, MA and FM discussed and interpreted the results. VSS provided the single crystals. All authors contributed to the manuscript. 

\subsection{Notes}
The authors declare that they have no competing financial interests.

\begin{acknowledgement}
The authors thank Y. Jin, Q. Dong and A. Cavanna for providing the cryogenic HEMT we used for the noise measurements, and M. Amato, H. Aubin and A. Mesaros for fruitful discussions. FM would like to acknowledge funding from H2020 Marie Sk\l{}odowska-Curie Actions (grant number 659247) and the ANR (ANR-16-ACHN-0018-01). The sample growth was supported by RSF-ANR grant number 20-42-09033.
\end{acknowledgement}

\bibliography{Single_atom_rotor}

\end{document}